\documentstyle[11pt]{article}

\newcommand{\be}{\begin{equation}}
\newcommand{\ee}{\end{equation}}
\newcommand{\bea}{\begin{array}}
\newcommand{\ea}{\end{array}}
\newcommand{\beqa}{\begin{eqnarray}}
\newcommand{\eeqa}{\end{eqnarray}}
\newcommand{\bean}{\begin{eqnarray*}}
\newcommand{\eean}{\end{eqnarray*}}
\newcommand{\eqn}[1]{(\ref{#1})}

\newcommand{\del}{\partial}

\def\up#1{\leavevmode \raise.16ex\hbox{#1}}

\newcommand{\journal}[4]{{\sl #1 }{\bf #2} \up(19#3\up) #4}

\newcommand{\gapproxeq}{\lower .7ex\hbox{$\;\stackrel{\textstyle >}{\sim}\;$}}
\newcommand{\lapproxeq}{\lower .7ex\hbox{$\;\stackrel{\textstyle <}{\sim}\;$}}

\newcounter{appendice}

\def\thebibliography#1{{\bf REFERENCES\markboth
 {REFERENCES}{REFERENCES}}\list
 {[\arabic{enumi}]}{\settowidth\labelwidth{[#1]}\leftmargin\labelwidth
 \advance\leftmargin\labelsep
 \usecounter{enumi}}
 \def\newblock{\hskip .11em plus .33em minus -.07em}
 \sloppy
 \sfcode`\.=1000\relax}

\begin{document}
\title{\hfill $\mbox{\small{
$\stackrel{\rm\textstyle DSF-33/97\quad}
{\rm\textstyle UAHEP 9711\quad}
{\rm\textstyle hep-th/9707153\quad\quad}
$}}$ \\[1truecm]
Hidden Quantum
Group Structure in\\ Einstein's General Relativity}
\author{G. Bimonte$^{a}$, R. Musto$^{a}$, A. Stern$^{b}$ and
P. Vitale $^{a}$}
\maketitle
\thispagestyle{empty}

\begin{center}
{\it a)  Dipartimento di Scienze Fisiche, Universit\`a di Napoli,\\
Mostra d'Oltremare, Pad.19, I-80125, Napoli, Italy; \\
INFN, Sezione di Napoli, Napoli, ITALY.\\
\small e-mail: \tt bimonte,musto,vitale@napoli.infn.it } \\
{\it b) Department of Physics, University of Alabama,\\
Tuscaloosa, AL 35487, USA.\\
\small e-mail: \tt astern@ua1vm.ua.edu }\\
\end{center}

\begin{abstract}

A new formal scheme is presented in which Einstein's classical theory of
General Relativity appears as the common, invariant sector of a
one-parameter family of different theories. This is achieved by
replacing the  Poincar\'e group of the ordinary tetrad formalism with a
$q$-deformed Poincar\'e group, the usual theory being recovered at
$q=1$. Although written in terms of {\it noncommuting} vierbein and
spin-connection fields, each theory has the same metric sector leading
to the ordinary Einstein-Hilbert action and to the corresponding
equations of motion. The Christoffel symbols and the components of the
Riemann tensor are ordinary {\it commuting} numbers and have the usual
form in terms of a metric tensor built as an appropriate bilinear in the
vierbeins. Furthermore we exhibit a one-parameter family of Hamiltonian
formalisms for general relativity, by showing that a canonical formalism
\`a la Ashtekar can be built for any value of $q$. The constraints are
still polynomial, but the Poisson brackets are {\it not} skewsymmetric
for $q\neq 1$.

\end{abstract}

\bigskip

{\bf PACS}:  04.20, 11.30.Cp, 11.15.-q

{\bf Keywords}: General Relativity, Hamiltonian Formalism, Quantum Groups

\newpage
\section*{Introduction}

Currently one of the most fascinating and challenging problems
in gravity involves understanding  the nature of space-time at
distance scales   where quantum mechanical effects enter.
 In the past decade, string theory has addressed this issue, but we
will not be concerned  with the string theory point of view here.
 Rather, we will be interested in approaching the problem in the
general framework of field theory. In this context,
a necessary step for   going  beyond the classical theory
is to pick the ``right set of variables" to construct a
canonical formalism suitable for quantization.
 This task is both
essential and nontrivial for a highly nonlinear theory such as general
relativity.  Here one may recall the success obtained in 2+1
Einstein's gravity from showing its equivalence with a Yang-Mills theory
described by a pure Chern-Simons action \cite{wit}.
In fact, the essential
progress allowed by the gauge theory formulation was to unravel the
structure of the classical phase space opening the way to the canonical
formalism and to quantization.

As is well known, it is not possible to formulate Einstein's gravity in
3+1 dimensions entirely as a Yang-Mills theory. Nevertheless, its
description in terms of variables such as
 vierbeins and spin connections, from which one constructs the metric
tensor and the Christoffel symbols,  has been a
fundamental step forward. In terms of these fields Einstein's gravity
appears as a gauge theory associated with the Poincar\'e group, although
the action only exhibits an invariance under the local Lorentz subgroup
(as well as under diffeomorphisms of the space-time manifold)
 \cite{weyl},\cite{uti}. Then gravitational interactions with matter are
prescribed by a gauge principle in analogy with all other fundamental
interactions as  dictated by the Standard Model.
Furthermore, following this approach, a much better understanding of the
structure of the classical phase space and great simplicity in building a
canonical formalism have been achieved. To be more specific, Ashtekar
\cite{ashtekar} has been able to construct a canonical formalism in which
the pull-backs on the ``space" manifold of the self-dual part of the
spin connections play the r\^ole of dynamical variables. One can then
identify the corresponding   conjugate momenta and show that
the constraints are polynomial  in  these variables.
However, a reality condition on physical solutions must be further
imposed.

There is still one more lesson that can be learned from $2+1$ dimensional
gravity. It has been recently shown \cite{bmsv} that the usual equivalence
between Einstein's theory and  Chern-Simons  theory with local Poincar\'e
group invariance is only a specific case of a more general equivalence.
Indeed Einstein's theory in 2+1 dimensions (in the absence of a
cosmological term) admits a one-parameter family of Chern-Simons
formulations, corresponding to the  $q$-deformed Poincar\'e gauge group
\cite{castcom}. Here $q$ is a real dimensionless parameter and the ordinary
Poincar\'e group, $ISO(2,1)$ is recovered for
 $q=1$.   For $q\ne 1$, the system has a noncommutative structure, which
we will elaborate on shortly.   The question which then naturally
arises is whether a similar noncommutative structure can also be present
in Einstein's gravity in
$3+1$ dimensions. In this paper we give a positive
answer to this question by showing that (torsionless) Einstein's gravity
 may be formulated as a  gauge theory associated with
 a $q$-deformed  Poincar\'e group.  The dynamics is determined from an
action analogous to
Palatini's, and it has the usual local Lorentz invariance.
We thus find that the usual description of Einstein's gravity
in terms of vierbeins and spin connections may be extended to a
one-parameter family of gauge theories.

It should be stressed that such an equivalence not only holds
for the pure gravity case, but it  also holds
in the presence of matter, provided there are
no sources for torsion.
(In fact, it is only a nonzero torsion that distinguishes the
different classical theories  from one another, each one
 coupling to a different kind of ``exotic" matter.)  As a result,
for zero torsion, our one-parameter  family of systems
 have the metric sector of the theory  in common.  It is the latter
 which  contains all the physically relevant information for
\underline{classical} gravity.  On the other hand, concerning
quantization, for each theory there exists
a canonical formalism \`a la Ashtekar.

Before advancing further in this discussion, a note of warning
should be made about
 the price paid for achieving these results.  It
concerns the noncommutative structure stated above.   A quantum group $G_q$
is defined as the \underline{noncommutative} algebra of functions on the
Lie-group $G$. As a consequence the gauge fields transforming locally under
a quantum (or $q$-deformed) group are \underline{ not} ordinary functions,
but exhibit nontrivial braiding relations among themselves
\cite{cast1,cast2}.
 (Let us stress that in our approach space-time is
an ordinary manifold labeled by commuting variables.) Vierbeins and
spin connections are fields, or more precisely
 differential one-forms, endowed  with nonstandard commutation relations.
As in the ordinary undeformed theory,
one can build a symmetric space-time
 metric as an appropriate  bilinear in the
vierbeins.  We find that
 different metric components commute among themselves, but do
\underline{not} commute with vierbeins and spin-connections. Nevertheless
the Christoffel symbols and the Riemann tensor are given by the usual
expression  in
terms of the metric tensor and its inverse, and furthermore, they
  commute with \underline{all} fields and therefore can be represented
  by ordinary numbers\footnote{For the case when matter is present,
 the same result holds for the energy momentum tensor.}. The
entire metric sector is made out of objects which mutually commute
and is identical to (torsionless) Einstein's gravity. The
noncommutative structure of the theory cannot then be probed with
large scale gravitational experiments where quantum effects are not
present.

The formal scheme which we propose,  where classical Einstein's
gravity appears as the common, invariant part of a one-parameter family of
gauge theories, may be quite interesting in itself, but
more interesting are its physical
consequences. As we already mentioned, for each of these theories, despite
the noncommutative nature of the variables, a canonical formalism \`a la
 Ashtekar can be carried out.  We find that the notion of self-duality is
consistent with the fields braiding relations and the constraints are still
polynomial, even if deformed with respect to the usual ones.   For
$q\ne 1$, the Poisson
brackets, however, have new features, such as not being
skewsymmetric due to the noncommuting nature of the conjugate variables.
It may be too early to say what may be the consequences of having a
family  of canonical formalisms for general relativity,
 even if expressed in terms of exotic
variables. However it seems fair to say that the
challenging problem of quantizing a deformed gauge theory may
be physically relevant.

We begin in section 1 by introducing the quantum Poincar\'e group.
This quantum group contains the (undeformed) Lorentz group.
In section 1, we also give a heuristic description
of a bicovariant differential calculus on the quantum group,
 which is necessary to
formulate the associated gauge theory.  The latter
  is done in section 2.
There we also write down an action principle for gravity
 which has local Lorentz invariance (as well as diffeomorphism
symmetry) and it turns out to be equivalent to the one found by Castellani
\cite{cast2}.
We next
show how to recover the metric theory, for the case of pure gravity
in section 3, and the case of coupling to matter (with no
torsion) in section 4.    We then apply Ashtekar's procedure in
section 5, and give concluding remarks in section 6.

\section{Bicovariant Differential Calculus on the Quantum Poincar\'e Group}
\setcounter{equation}{0}
In this section we give a heuristic description
 of a quantum Poincar\'e group,  which we denote by
$ISO_q(3,1)$, together with a bicovariant differential calculus on it. The
procedure is the same as the one followed in \cite{bmsv} for the case of
$2+1$ dimensions.  We refer the reader to
 it for more details.  The
 mathematical structures discussed here are equivalent to those found in
\cite{castcom}, where we refer the reader for rigorous proofs.

We begin with  the 3+1 Poincar\'e group $ISO(3,1)$.  It
is most easily described in terms of
Lorentz matrices $\ell = [\ell_{ab}]$ and vectors $z= [z_a]$.
 Roman letters from
the beginning of the alphabet, running from 1  to 4, denote Lorentz
indices.  We shall raise and lower them using the Lorentz metric tensor
\begin{equation}
\eta=\pmatrix{ & & & 1\cr & 1 & & \cr & & 1 & \cr 1 & & & \cr}\;.
\label{4dmet}
\end{equation}
 Infinitesimal left and right transformations on the group are given
by the  variations
\begin{equation}
\delta_L {\ell}^{ba}= \tau_L^{ bc} {\ell_c}^{a} \;,\qquad \delta_L z^b =
\tau_L^ {bc} \; z_c + \rho_L^b \;, \label{ltr}
\end{equation}
and
\begin{equation}
\delta_R {\ell_c}^{d}= \ell_{cf}\;\tau_R^{fd} \;,\qquad \delta_R z^b = {
\ell^b}_a \;\rho_{R}^a \;, \label{rtr}
\end{equation}
respectively, with $\tau_L^{ab} = -\tau_L^{ba}$ and $\tau_R^{ab} =
-\tau_R^{ba}$ being  infinitesimal Lorentz parameters and
$\rho_{L}^a$ and  $\rho_{R}^a$   being infinitesimal translations.

The quantum Poincar\'e group $ISO_q(3,1)\equiv Fun_q(ISO(3,1))$
can also be described in terms of matrix
elements $\ell_{ab}$ and vector components $z_{a}$,
 but, unlike for  $ISO(3,1)$, they are
not c-numbers.  Instead they obey the following commutation relations:
\begin{equation}
z^a \;{\ell_c}^{b} = q^{\Delta(b)} \; {\ell_c}^{b} \;z^a \;,\label{4dcz}
\end{equation}
where
\begin{equation}
\Delta(1)=-1 \;,\quad \Delta(2)=\Delta(3)=0 \;,\quad \Delta(4)=1 \;,
\label{Delts}\end{equation}
and all other commutation relations are trivial. These commutation
relations are associated with an $R$-matrix satisfying the quantum
Yang-Baxter equation \cite{castcom}. They are consistent with the Lorentz
constraint
\be
\ell_{ab} \ell_c^{~b} = \eta_{ac}~,\label {dolm}
\ee
due to the identity
\begin{equation}
\eta_{ab}= q^{\Delta(a)+\Delta(b)} \; \eta_{ab} \;, \label{eteqet}
\end{equation}
which follows from the metric  (\ref{4dmet}) along with (\ref{Delts}).
   As in \cite{bmsv}, the commutation relations (\ref{4dcz})
are also consistent with the other constraints
defining the $SO(3,1)$ group, namely
$\ell_{ba} \ell^b_{~c} = \eta_{ac}$ and det$(\ell)=1$.  After imposing
all such constraints we therefore conclude that $ISO_q(3,1)$ contains
the ordinary Lorentz group.

Since $\ell_{ab}$ and $z_a$ are not
$c$-numbers, neither are the infinitesimal transformation parameters
 $\tau_L^{ab}$, $\tau_R^{ab}$, $\rho_{L}^a$ and  $\rho_{R}^a$.
 In fact, for the  commutation
relations \eqn{4dcz} to be preserved, we need
\beqa
\rho_L^a \;{\ell^c}_{b} & =& q^{\Delta(b)} \; {\ell^c}_{b}\;\rho_L^a \cr
\rho_L^a\; z^{b} & =& z^{b}  \rho_L^a     \label{lztrL}
\eeqa
and
\beqa
\rho_R^a \;{\ell_c}^{b} & =& q^{\Delta(b)}\; {\ell_c}^{b}\;\rho_R^a \cr
\rho_R^a\; z^{b} & =& q^{-\Delta(a)} \; z^{b}\; \rho_R^a \cr
\tau_R^{ab}\; z^c & =& q^{-\Delta(a)-\Delta(b)} \;
z^{c}\;\tau_R^{ab}\;. \label{rpar}
\eeqa
Also, the Lorentz transformation parameters $\tau_L$ commute with
${\ell}_c^{\;b}$ and $z^a$, while $\tau_R$ commute with
 ${\ell}_{c}^{\;b} $.

In order to write a differential calculus on the space spanned by $\ell_{ab}$
and $z_{a}$, we must specify the commutation rules among $l_{ab}, z_a$ and
their exterior derivatives.  A natural choice, consistent with
\eqn{4dcz} is
\begin{eqnarray}
d z^a \;{\ell_c}^{b} & = & q^{\Delta(b)} \; {\ell_c}^{b} \;dz^a \cr z^a \;d{
\ell_c}^{b} & = & q^{\Delta(b)} \;d {\ell_c}^{b} \;z^a \cr d z^a \wedge d{
\ell_c}^{b} & = & - q^{\Delta(b)} \;d {\ell_c}^{b} \wedge d z^a \;,
\label{dzel}
\end{eqnarray}
while we assume the calculus on the space spanned by $l_{ab}$ alone to be
the usual one on $SO(3,1)$
\beqa
d{ {\ell}^b}_a\;{ {\ell}_c}^d & = & {\ell_c}^{d} \;d{{\ell}^b}_a \cr
d{ {\ell}^b}_a\;\wedge d{{\ell}^d}_c & =& -d{\ell^d}_c \;\wedge
d{{\ell}^b}_a  \;, \label{ll}
\eeqa
and the calculus on the space generated by $z_a$ alone to be the usual
one on ${\bf R}^4$
\beqa
d z^b\; z^d & = & z^{d} \;dz^b \cr
d z^b\; \wedge dz^d & = & -dz^{d} \wedge \;dz^b   \;.
\label{zz}   \eeqa
In addition to these commutation relations,
we assume that the commutation relations
(\ref{lztrL}) and (\ref{rpar}) also hold when we replace $\ell$ and $z$
by their exterior derivatives.

The bicovariant bimodule of one--forms on the group is spanned either by
left or by right invariant one--forms.  We choose to work with the left
invariant basis and denote the one forms by
$\omega ^{ab} = - \omega ^{ba}$  and  $ e ^c $.    The following
expressions are invariant under (global)  left transformations:
\begin{equation}
\omega ^{ab} = (\ell^{-1} d\ell)^{ab} \;,\quad e ^c
=(\ell^{-1} dz)^{c} \;.  \label{4dli}
\end{equation}       On the other hand,
 under infinitesimal right transformations
(\ref{rtr}) they undergo the variations
\begin{eqnarray}
\delta_R \omega^{ab} & =& d\tau_R^{ab} + {\omega^a}_c \;\tau_R^{cb} -
{\omega^b}_c \;\tau_R^{ca} \;,\cr
\delta_R e^c & = & d\rho^c_R +
{\omega^{c}}_b \;\rho^b_R - \tau_R^{cb}\;e_b\;.  \label{4dtras}
\end{eqnarray}
~From (\ref{rpar}) we get the following commutation relations between
 gauge parameters and one--forms
\begin{eqnarray}
\rho_R^a \;\omega^{bc} & =& q^{\Delta(b)+\Delta(c) } \;\omega^{bc}\;
\rho_R^a \cr
\rho_R^a\; e^{b} & =& q^{\Delta(b)-\Delta(a)} \; e^{b}\;  \rho_R^a \cr
\tau_R^{ab} \; e^c & =& q^{-\Delta(a)-\Delta(b)} \; e^{c} \;
\tau_R^{ab}\;~ \cr
\tau_R^ {ab} \omega^{cd} &=& \omega^{cd}  \tau_R^ {ab} \;. \label{crgpc}
\end{eqnarray}
The left invariant forms (\ref{4dli}) satisfy the Maurer Cartan equations
\beqa
{\cal R}^{ab}&=& 0~,\label{ceq1}  \\
{\cal T}^a& =&0 \;,  \label{ceq2}
\eeqa
where ${\cal R}^{ab}$ and ${\cal T}^a $ have the usual expressions for the
curvature and torsion
\begin{eqnarray}
{\cal R}^{ab} & =& d\omega^{ab}+ {\omega^a}_c\wedge \omega^{cb}\;, \cr
{\cal T}^a & =& de^a+ {\omega^a}_b\wedge e^b\;,  \label{4dRT}
\end{eqnarray}
except that we no longer have the usual exterior product for the one forms $
\omega^{ab}$ and $e^a$. From (\ref{ll}), (\ref{zz}),  (\ref{dzel}) and
(\ref{4dli}) we instead get
\begin{eqnarray}
\omega^{ab}\wedge \omega^{cd} & =& - \omega^{cd}\wedge \omega^{ab} \;,\cr
e^a\wedge \omega^{bc} & =& - q^{\Delta(b) +\Delta(c)} \; \omega^{bc}\wedge
e^a \;,\cr e^a\wedge e^b & =& - q^{\Delta(b) -\Delta(a)} \; e^b\wedge e^a
\;.
\label{4daa2}
\end{eqnarray}

\section{One Parameter Family of Tetrad Theories}
\setcounter{equation}{0}
Using the mathematical framework of the previous section, we shall construct
a general q-Poincar\'e gauge  theory and then write down an action
principle for gravity having the usual symmetries, i.e.
diffeomorphism invariance and local Lorentz invariance.

A general q-Poincar\'e gauge  theory is obtained when we drop the
assumption   (\ref{4dli}) that
$\omega ^{ab} $ and $e^a$ be pure gauges, and hence that they
satisfy the Maurer Cartan equations (\ref{ceq1}) and (\ref{ceq2}).
 Rather, we let them be
arbitrary spin connections and vierbein one forms.  In this case,
the right  transformations  (\ref{4dtras}) are to be regarded as
infinitesimal gauge transformations.  Such transformations on the
curvature and torsion are given by
\begin{eqnarray}
\delta_R {\cal R}^{ab} & =& {{\cal R}^a}_c \;\tau_R^{cb}-{{\cal R}^b}_c
\;\tau_R^{ca} \;,\cr \delta_R {\cal T}^c & = &{{\cal R}^{c}}_b \;\rho^b_R
- \tau_R^{cb}\;{\cal T}_b\;.  \label{gtoRT}
\end{eqnarray}
 The Bianchi identities for this theory take the usual form
\begin{eqnarray}
d{\cal R}^{ab} & =& {\ {\cal R}^a}_c\wedge \omega^{cb}- {\ {\cal R}^b}
_c\wedge \omega^{ca}\;, \cr d{\cal T}^a & =& {{\cal R}^a}_b\wedge e^b - {
\omega^a}_b\wedge {\cal T}^b \;,  \label{4dBi}
\end{eqnarray}
 the ordering being crucial.  We assume, consistent with
\eqn{4dRT} and \eqn{4daa2}, that
\beqa
\omega^{ab}\wedge {\cal R}^{cd} &=& {\cal R}^{cd}\wedge \omega^{ab}\;,\cr
\omega^{bc}\wedge {\cal T}^a & =& q^{-\Delta(b) -\Delta(c)} \; {\cal T}
^a\wedge \omega^{bc} \;,\cr e^a\wedge {\cal R}^{bc} & =& q^{\Delta(b)
+\Delta(c)} \; {\cal R}^{bc}\wedge e^a \;,\cr e^a\wedge {\cal T}^b &
=&q^{\Delta(b)-\Delta(a)}\;{\cal T}^b\wedge e^a \;,  \label{4dRToe}
\eeqa
and
\beqa
{\cal R}^{ab}\wedge {\cal R}^{cd} & =& {\cal R}^{cd}\wedge {\cal R}^{ab} \;,
\cr
{\cal R}^{bc}\wedge {\cal T}^a & =& q^{-\Delta(b) -\Delta(c)} \; {\cal T}
^a\wedge {\cal R}^{bc} \;,\cr
{\cal T}^a\wedge {\cal T}^b & =& q^{\Delta(b)
-\Delta(a)} \; {\cal T} ^b\wedge {\cal T} ^a \;.  \label{4dRTRT}
\eeqa
With a rescaling of the spin connection $\omega^{ab}$ and spin
curvature ${\cal R}^{ab}$  by a factor of
$q^{-{1\over 2}(\Delta(a) + \Delta(b))}$, the above system becomes
identical to that of ref. \cite{cast2}.

The above gauge theory may be associated with a q-Lie algebra,
defined by the algebraic relations \beqa
T_i T_j - \Lambda_{ij}^{kl} \, T_k T_l=C_{ij}^k T_k \;,\label{TiTj}\eeqa
 where $T_i$ are
the generators, $ \Lambda_{ij}^{kl}$ the braiding matrix elements and
$C_{ij}^k $ are the q-structure constants.  Following \cite{cast1}, from
the commutation relations  (\ref{crgpc})
 we can identify the braiding matrix, while from the gauge
 transformations   (\ref{4dtras})   we can identify
the q-structure constants.  For the former we get
\begin{equation}
\Lambda_{ab\;cd}^{cd\;ab}=1 \;,\quad
\Lambda_{a\;\;bc}^{bc\;\;a}=q^{-\Delta(b)-\Delta(c)} \;,\quad
\Lambda_{bc\;a}^{a\;bc}=q^{\Delta(b)+\Delta(c)} \;,\quad
\Lambda_{a\;b}^{b\;a}= q^{\Delta(a)- \Delta(b)} \;,
\end{equation}
with all other components vanishing.
It is then  verified that the square of the braiding matrix is the
unit matrix, and furthermore that
 $ \Lambda_{ij}^{kl}$ and
$C_{ij}^k $ satisfy all the necessary conditions (see \cite{cast1})
of a minimally deformed gauge theory.     The ten generators $T_i$,
which we denote by $M_{ab}$ (Lorentz generators) and $P_a$ (translations),
are said to be dual to the one
forms  $\omega^{ab}$ and $e^a$.
In terms of them  (\ref{TiTj}) is expanded to
\beqa
&&[M_{ab}, M_{cd}] = \eta_{ac} M_{bd} - \eta_{bc} M_{ad} +
\eta_{bd} M_{ac} - \eta_{ad} M_{bc} \cr
&&[M_{ab}, P_c]_{q^{\Delta (a)+ \Delta (b)}} = -(\eta_{bc} P_{a}
- \eta_{ac} P_{b} )\cr
&&[P_a, P_b]_{q^{\Delta (a)- \Delta (b)}} = 0~, \label{qlie}
 \eeqa      where $[\alpha,\beta]_s\equiv\alpha\beta - s\beta\alpha$.

Next we write down  a locally Lorentz invariant action:
\be
{\cal S} =\frac{1}{4} \int_M \; \epsilon_{abcd} {\cal R}^{ab} \wedge {\cal
E}^{cd}
\;,
\label{4dact}
\ee
where ${\cal E}^{cd} $ is the two form
\be
{\cal E}^{cd}= -{\cal E}^{dc} = q^{-\Delta(d)} e^c \wedge e^d \;,
\ee
$M$ is a four manifold and $\epsilon_{abcd} $ is the ordinary, totally
antisymmetric tensor with $\epsilon_{1234} =1$. The expression
(\ref{4dact}) differs from that of  the undeformed case by the
$q^{-\Delta(d)} $ factor. Note that this factor can be written
differently using the identity
\be
q^{\Delta(a)+\Delta(b)+ \Delta(c)+ \Delta(d)} \; \epsilon_{abcd}
=\epsilon_{abcd}\;\;.  \label{epsid}
\ee
It can be checked that the action \eqn{4dact} is identical to the
 one found in \cite{cast2}, up to an overall factor $q^{3/2}$.
To check local Lorentz invariance we use the property
\begin{equation}
e^a\wedge \delta e^b=-q^{\Delta(b) -\Delta(a)} \;\delta e^b\wedge e^a \;.
\end{equation}
Then
\begin{equation}
\delta {\cal S}=\frac{1}{4}\int_M q^{-\Delta(d)}\; \epsilon_{abcd}(\delta {\cal
 R}
^{ab} \wedge e^c + 2{\cal R}^{ab} \wedge \delta e^c)\wedge e^d \;.
\end{equation}
After substituting in the variations (\ref{4dtras}) and (\ref{gtoRT}) with $
\rho^b_R = 0$ we get
\begin{equation}
\delta {\cal S} =\frac{1}{2} \int_M \; \epsilon_{abcd} ({{\cal R}^{a}}_f\wedge
\tau^{fb}_R {\cal E}^{cd} - {\cal R}^{ab} \wedge \tau^{cf}_R {{\cal E}_f}^d) \;,
\end{equation}
which vanishes due to the antisymmetry of $\tau^{fb}_R$ and ${\cal
E}^{cd}$. Also, as in the undeformed case, the action is invariant
 under the full set of local Poincar\'e
 transformations (\ref{4dtras}), provided we impose the
torsion to be zero upon making the variations.

The equations of motion obtained from varying the vierbeins have the usual
form, i.e.
\begin{equation}
\; \epsilon_{abcd} {\cal R}^{ab} \wedge e^{c} = 0 \;,  \label{ffeq1}
\end{equation}
while varying $\omega^{ab}$ gives
\begin{equation}
d\tilde {\cal E}_{ab} = {\omega_a}^c \tilde {\cal E}_{bc} - {\omega_b}^c
\tilde {\cal E}_{ac}
\;,\qquad \tilde {\cal E}_{ab}\equiv \epsilon_{abcd} {\cal E}^{cd} \;.
 \label{eomfcE}
\end{equation}
Due to the antisymmetry of ${\cal E}^{cd}$, we get the following
expression in
terms of the torsion from  (\ref{eomfcE})
\begin{equation}
\; \epsilon_{abcd} {\cal T}^{c} \wedge e^{d}\; q^{-\Delta(d)} = 0 \;.
\label{ffeq2}
\end{equation}    In the next section we show that this equation
implies zero torsion (\ref{ceq2}), provided inverse
vierbeins exist. This is necessary in order to recover Einstein's gravity.

\section{Recovering Einstein's theory}
\setcounter{equation}{0}
In this Section we prove that the metric formulation of the q-deformed
Cartan theory of gravity discussed in the previous Section is completely
equivalent to the {\it undeformed} Einstein's theory, for all values of
$q$.

To make a connection with Einstein gravity, we need to introduce the
space-time metric ${\tt g}_{\mu
\nu}$ on $M$. In order to do it we assume the following stronger form of
the commutation relations
\eqn{4daa2}:
\begin{eqnarray}
\omega _\mu ^{ab}\omega _\nu ^{cd}=\omega _\nu ^{cd}\omega _\mu ^{ab}\;,\cr
e_\mu ^a\omega _\nu ^{bc}=q^{\Delta (b)+\Delta (c)}\;\omega _\nu ^{bc}e_\mu
^a\;,\cr e_\mu ^ae_\nu ^b=q^{\Delta (b)-\Delta (a)}\;e_\nu ^be_\mu ^a\;,
\label{4dcrstc}
\end{eqnarray}
$e_\mu ^a$ denoting the space-time components of the vierbein one form
$e^a$ ; $\mu $ and $\nu $ being space-time indices.   These commutation
relations are consistent with the view that the space-time manifold
be spanned by commuting coordinates.    We now want  to
construct a bilinear from
 the vierbeins which is symmetric in the space-time
indices and invariant under local Lorentz transformations. These
requirements uniquely fix ${\tt g}_{\mu\nu}$ to be
\begin{equation}
{\tt g}_{\mu \nu }=q^{\Delta (a)}\;\eta _{ab}\;e_\mu ^ae_\nu ^b\;,
\label{4dsymmet}
\end{equation}
Using eqs.\eqn{4dcrstc} we see that ${\tt g}_{\mu
\nu }$ is symmetric, although the tensor elements are not c-numbers since
\beqa
{\tt g}_{\mu \nu }\;\omega _\rho ^{ab}=q^{2\Delta (a)+2\Delta (b)}\;
\omega_\rho ^{ab}\;{\tt g}_{\mu \nu }\;,\cr {\tt g}_{\mu \nu }\;e_\rho
^a=q^{2\Delta (a)}\;e_\rho ^a\;{\tt g}_{\mu \nu }\;.
\eeqa

The components of ${\tt g}_{\mu \nu }$ do however commute with themselves.

In order to go further, we need be able to define the inverses $e^{\mu}_a$
of the (co-)tetrads $e^a_{\mu}$. This can be done if we enlarge our algebra
by a new element ${\tt e}^{-1}$ such that:
$$
{\tt e}^{-1} e^a_{\mu}= q^{-4 \Delta(a)}e^a_{\mu} \; {\tt e}^{-1}~,
$$
\begin{equation}
{\tt e}^{-1} \omega^{ab}_{\mu}= q^{-4 (\Delta(a)+ \Delta(b))
}\omega^{ab}_{\mu}\; {\tt e}^{-1}~,\label{creminus}
\end{equation}
and such that
\begin{equation}
{\tt e}^{-1} {\tt e} =1~,\label{eeminus}
\end{equation}
where ${\tt e}$ is the determinant:
\be
{\tt e}=\epsilon^{\mu \nu \rho \sigma} e^1_{\mu} e^2_{\nu} e^3_{\rho}
e^4_{\sigma}~.
\ee
Eq.\eqn{eeminus} is consistent because its left hand side
 commutes with everything,
due to eqs.\eqn{creminus}. Moreover, one can check that ${\tt e}^{-1} {\tt e}
= {\tt e } {\tt e}^{-1}$.
The inverses of the vierbeins can now be written:
\begin{equation}
e^{\mu}_a= \frac{1}{3!} \hat{\epsilon}_{abcd} \epsilon^{\mu \nu \rho
\sigma} e^b_{\nu} e^c_{\rho} e^d_{\sigma} {\tt e}^{-1}~, \label{definv}
\end{equation}
where the totally q-antisymmetric tensor  $\hat{\epsilon}_{abcd} $
 is defined such that
\begin{equation}
\hat{\epsilon}_{abcd} e^a \wedge e^b \wedge e^c \wedge e^d = e^1 \wedge e^2
\wedge e^3 \wedge e^4 ~~~{\rm no~sum~on}~a,b,c,d
\end{equation}    The solution to this equation can be expressed by
\begin{equation}
\hat{\epsilon}_{abcd} = q^{3\Delta(a)+2\Delta(b)+\Delta(c) +3}\;
\epsilon_{abcd}  \;.\label{ephaep}\end{equation}
Notice also the following useful identity satisfied by the q-antisymmetric
tensor $\hat{\epsilon}^{abcd}$ obtained by raising the indices of
$\hat{\epsilon}_{abcd}$ with the metric $\eta^{ab}$
\be q^{-6}\; \hat{\epsilon}^{abcd}\;{\tt e}
=-\epsilon^{\mu\nu\lambda\sigma} e^a_\mu e^b_\nu e^c_\lambda e^d_\sigma
\;.\label{epep} \ee
The explicit expression of $\hat{\epsilon}^{abcd}$ can be seen to
be:
\begin{equation}
\hat{\epsilon}^{abcd} = q^{-3\Delta(a)-2\Delta(b)-\Delta(c) +3}\;
\epsilon^{abcd}  \;,\label{defeup} \end{equation}
where $\epsilon^{abcd}$ is the ordinary antisymmetric tensor obtained by
raising the indices of $\epsilon_{abcd}$ with the metric $\eta^{ab}$. It is
easy to prove that the inverses of the vierbeins
\eqn{definv} have the usual properties:
$$
e^a_{\mu} e_b^{\mu}=e_b^{\mu} e^a_{\mu} = \delta^a_b~,
$$
\begin{equation}
e^a_{\mu} e_a^{\nu} = e_a^{\nu} e^a_{\mu} = \delta^{\nu}_{\mu}~.
\label{prode}
\end{equation}

By using the inverses of the tetrads, we can now prove that eq.\eqn{ffeq2}
implies the vanishing of the torsion. To begin with, we introduce the
components of the torsion two-form along the tetrads:
\be
{\cal T}^a_{bc} \; \equiv \; q^{\Delta(b)} {\cal T}^a_{\mu \nu} e^{\mu}_b
e^{\nu}_c~;
\ee
the power of $q$ ensures that they are antisymmetric in the lower indices,
${\cal T}^a_{bc}=-{\cal T}^a_{cb}$. Now we rewrite eq.\eqn{ffeq2} as:
\begin{eqnarray}
0&=&q^{-\Delta(d)}\; \epsilon_{abcd} \epsilon^{\lambda \mu \nu
\rho} {\cal T}^c_{\mu \nu}e^d_{\rho}=
q^{-\Delta(d)-\Delta(h)}\; \epsilon_{abcd} \epsilon^{\lambda \mu \nu
\rho} {\cal T}^c_{gh} e^g_{\mu} e^h_{\nu} e^d_{\rho}=\nonumber\\
&=& - q^{\Delta(d) - \Delta(f)-3} \epsilon_{abcd}
\epsilon^{fghd} {\cal T}^c_{gh} e^{\lambda}_f {\tt e}=\nonumber\\
&=&2 q^{- \Delta(a) -\Delta(b) -\Delta(c) -3}(q^{-\Delta(a)}{\cal T}^c_{bc}
e^{\lambda}_a + q^{-\Delta(b)}{\cal T}^c_{ca} e^{\lambda}_b +
q^{-\Delta(c)}{\cal T}^c_{ab} e^{\lambda}_c)\;{\tt e}~,\label{torzero}
\end{eqnarray}
where we have used the identity
\be
q^{-\Delta(d)-\Delta(h)}\epsilon^{\lambda \mu \nu \rho}e^g_{\mu} e^h_{\nu}
e^d_{\rho}= -q^{\Delta(d)-\Delta(f)-3} \epsilon^{fghd} e^{\lambda}_f {\tt
e}~,
\ee
which follows from eqs.\eqn{epep} and \eqn{defeup}. Neglecting the overall
factor of $q^{-\Delta(a)
-\Delta(b)
-3}{\tt e}$ in
\eqn{torzero} and multiplying it on the right by $e^d_{\lambda}$ we finally
get
\be
q^{-\Delta(c)}{\cal T}^c_{bc} \; \delta^d_a + q^{-\Delta(c)}{\cal T}^c_{ca}
\; \delta^d_b +q^{- \Delta(d)} {\cal T}^d_{ab}=0~.
\ee
It is easy to verify that these equations imply the vanishing of all the
${\cal T}^a_{bc}$ and thus of the torsion.

We now define the Christoffel symbols $\Gamma
_{\mu
\nu }^\sigma $ by demanding that the covariant derivative of the vierbeins
vanishes,
\be
{\tt D}_\mu e_\nu ^b=0~,\label{covd}
\ee where
\begin{equation}
{\tt D}_\mu e_\nu ^b={\cal D}_\mu e_\nu ^b-\Gamma _{\mu \nu }^\sigma
e_\sigma ^b\;,  \label{cddr}
\end{equation}
and
\begin{equation}
{\cal D}_\mu e_\nu ^b=\partial _\mu e_\nu ^b+\omega _\mu ^{bc}e_{\nu c}\;.
\end{equation}
The torsion being zero is consistent with the Christoffel symbols being
symmetric in the lower two indices.

We can eliminate the spin connections from (\ref{covd}), if we multiply it
on the left by $q^{\Delta (a)}\eta
_{ab}e_\rho ^a$, sum over the $b$ index, and symmetrize with respect to the
space-time indices
 $\nu $ and   $\rho $. The result is
\begin{equation}
0=q^{\Delta (a)}\eta _{ab}[e_\rho ^a\partial _\mu e_\nu ^b+e_\nu ^a\partial
_\mu e_\rho ^b-e_\rho ^ae_\sigma ^b\Gamma _{\mu \nu }^\sigma -e_\nu
^ae_\sigma ^b\Gamma _{\mu \rho }^\sigma ]\;.
\end{equation}
Next we add to this the equation obtained by switching $\mu $ and $\nu $,
and subtract the equation obtained by replacing indices ($\mu ,\nu ,\rho )$
by ($\rho ,\mu ,\nu )$. We can then isolate $\Gamma _{\mu \nu }^\sigma $
according to
\begin{equation}
2q^{\Delta (a)}\eta _{ab}e_\rho ^ae_\sigma ^b\Gamma _{\mu \nu }^\sigma
=q^{\Delta (a)}\eta _{ab}[e_\rho ^a(\partial _\mu e_\nu ^b+\partial _\nu
e_\mu ^b)+e_\nu ^a(\partial _\mu e_\rho ^b-\partial _\rho e_\mu ^b)+e_\mu
^a(\partial _\nu e_\rho ^b-\partial _\rho e_\nu ^b)]
\end{equation}
or
\begin{equation}
2{\tt g}_{\rho \sigma }\Gamma _{\mu \nu }^\sigma =\partial _\mu {\tt g}
_{\rho \nu }+\partial _\nu {\tt g}_{\rho \mu }-\partial _\rho {\tt g}_{\nu
\mu }\;.  \label{chris}
\end{equation}

To solve this equation we need the inverse of the metric ${\tt g}^{\mu
\nu}$. The expression
\begin{equation}
{\tt g}^{\mu \nu}= q^{\Delta(a)} \eta^{ab} e^{\mu}_a e^{\nu}_b  \;,
\end{equation}
does the job as it can be checked that
\begin{equation}
{\tt g}^{\mu \rho} {\tt g}_{\rho \nu}=
{\tt g}_{\nu \rho} {\tt g}^{\rho \mu}= \delta^{\mu}_{\nu} \;.
\end{equation}
Notice that  unlike in the usual Einstein Cartan theory
\be
  {\tt g}^{\mu \nu} \eta_{ab} e^b_{\nu}= q^{\Delta(a)} e^{\mu}_a~.\label{unl}
\ee

 We are now able to solve eq.\eqn{chris}.  Upon multiplying it by ${\tt
g}^{\tau \rho}$ on any side, we get the usual expression for the
Christoffel symbols in terms of the metric tensor and its inverse. It
may be verified, using  these expressions, that the Christoffel symbols
commute with everything and thus, even if written in  terms of
non-commuting quantities, they can be interpreted as being ordinary
numbers.

The covariant derivative operator $\nabla_{\mu}$ defined by the Christoffel
symbols is compatible with the metric ${\tt g}_{\mu \nu}$, i.e.
 $\nabla_{\mu} {\tt g}_{\nu
\rho}=0$. This is clear because our Christoffel symbols have
the standard expression in terms of the space-time metric ${\tt g}_{\mu
\nu}$, but also follows from eq.\eqn{covd}
\be
\nabla_{\mu} {\tt g}_{\nu \rho}={\tt D}_{\mu} {\tt g}_{\nu \rho}=
{\tt D}_{\mu}( q^{\Delta (a)} \eta_{ab} e^a_{\nu} e^b_{\rho})=0~.
\ee
We now construct the Riemann tensor. It is defined as in the undeformed
theory:
\be
{\tt R}_{\mu \nu \rho}^{~~~~\sigma} v_{\sigma} =({\tt D}_{\mu} {\tt
D}_{\nu}-{\tt D}_{\nu} {\tt D}_{\mu} ) v_{\rho}~,
\ee
where $v_{\mu}$ is an ordinary co-vector. It follows from \eqn{covd} that
it has the standard expression in terms of the Christoffel symbols (and
thus in terms of the space-time metric and its inverse) and therefore its
components commute with everything. (This is also true for the Ricci tensor
${\tt R}_{\mu \nu}={\tt R}_{\mu \sigma
\nu}^{~~~~\sigma}$, of course, but not for ${\tt R}_{\mu \nu \rho \tau}$
as the lowering of the upper index of the Riemann tensor implies
contraction with ${\tt g}_{\sigma \tau}$ which is not in the center of the
algebra). The relation among the Riemann tensor and the curvature of the
spin connection follows from eqs.\eqn{covd} and \eqn{cddr}:
\be
e_{\sigma}^a {\tt R}_{\mu \nu \rho}^{~~~~\sigma} v^{\rho}
=e_{\sigma}^a({\tt D}_{\nu} {\tt D}_{\mu}- {\tt D}_{\mu} {\tt D}_{\nu})
 v^{\sigma}=({\cal
D}_{\nu} {\cal D}_{\mu}- {\cal D}_{\mu} {\cal D}_{\nu})e_{\sigma}^a
v^{\sigma}= -{\cal R}_{\mu
\nu }^{ac}\eta_{bc} e_{\sigma}^b v^{\sigma}~,
\ee
$ {\cal R}^{ab}_{\mu \nu}$ being the space-time components of $ {\cal
R}^{ab}$. $v^{\mu}$ being an arbitrary ordinary vector, it follows from the
above equation that:
\be
{\tt R}_{\mu \nu \rho}^{~~~~\tau}=-{\cal R}^{ac}_{\mu \nu }\eta_{bc}
e^b_{\rho}e^{\tau}_a~.\label{riecur}
\ee
Using this equation it can be checked directly that the components of the
Riemann tensor commute with everything, as pointed out earlier. Our Riemann
tensor has the usual symmetry properties:
\begin{eqnarray}
{\tt R}_{\mu \nu \rho}^{~~~~\sigma}& = & -{\tt R}_{\nu \mu
\rho}^{~~~~\sigma}~,\nonumber\\{\tt R}_{\mu \nu
\rho \sigma}& =& - {\tt R}_{\mu
\nu \sigma \rho}
~,\nonumber\\{\tt R}_{[\mu \nu \rho]}^{~~~~~\sigma}& = & 0~. \label{riem}
\end{eqnarray}
The first of these equations is obvious; the second can be proved starting
from \eqn{riecur}:
\begin{eqnarray}
&{\tt R}_{\mu \nu \rho \sigma}&={\tt R}_{\mu \nu \rho}^{~~~~\tau} {\tt
g}_{\tau
\sigma}=
-{\cal R} ^{ab}_{\mu \nu} e_{\rho b}e^{\tau}_a {\tt g}_{\tau \sigma}=\nonumber\\
&=&-q^{\Delta(a)}{\cal R} ^{ab}_{\mu \nu}e_{\rho b} e_{\sigma a} =
-q^{\Delta(b)}{\cal R} ^{ab}_{\mu \nu}e_{\sigma a} e_{\rho b}=
-{\tt R}_{\mu \nu \sigma \rho}~,
\end{eqnarray}
where we have made use of \eqn{unl}. The third of eqs.\eqn{riem} follows
from the algebraic Bianchi identity, namely the second of eqs.\eqn{4dBi}
with ${\cal T}^a=0$, and from \eqn{riecur}:
\be
0=-\epsilon^{\lambda \mu \nu \rho}{\cal R}^{ac}_{\mu \nu } \eta_{bc}
e^b_{\rho}=\epsilon^{\lambda \mu \nu \rho}{{\tt R}_{\mu \nu
\sigma}}^{\tau} e^a_{\tau} e^{\sigma}_b e^b_{\rho}=6 \;
\epsilon^{\lambda \mu \nu
\rho}{\tt R}_{[\mu
\nu \rho]}^{~~~~\tau} e^a_{\tau}~.
\ee

We now show that the action \eqn{4dact} becomes equal to the {\it
undeformed} Einstein-Hilbert action, once the spin connection is eliminated
using its equations of motion, namely the zero torsion condition. For the
purposes of the next Sections we shall do it in two steps: first we rewrite
\eqn{4dact} in a form analogous to Palatini's action, which we shall use to
 develop
the canonical formalism, and then show that the latter reduces to the {\it
undeformed} Einstein-Hilbert action, once the spin-connection is eliminated
from it. Consider thus the following deformation of the Palatini action:
\begin{equation}
{\cal S}=\frac{1}{2}\int_M d^4 x \, q^{\Delta(a)-3} {\tt e} \, e^{\mu}_a
e^{\nu}_b {\cal R}^{ab}_{\mu \nu}~.
\label{qpala}
\end{equation}
To see that it coincides with \eqn{4dact}, we use the identity:
\be q^{\Delta(a)-\Delta(b)-6}    \hat{\epsilon}^{abcd}
 e^\mu_a e^\nu_b {\tt e}   =
-\epsilon^{\mu\nu\lambda\sigma} e^c_\lambda e^d_\sigma \;. \label{id68}
\ee  The
result (\ref{qpala}) then follows after multiplying both sides of this
equation on the left by $-1/8\; q^{-2\Delta(f) - \Delta(g)-3}
\hat{\epsilon}_{fgcd} {\cal R}^{fg}_{\mu\nu} $ and using the identity
$$\hat{\epsilon}_{fgcd}
 \hat{\epsilon}^{abcd} =-2 q^6\;(\delta^a_f\delta^b_g-q^{\Delta(f)-\Delta(g)}
\delta^a_g\delta^b_f)\;,$$ along with (\ref{ephaep}).

We now show that eq.\eqn{qpala} becomes in turn equal to the undeformed
Einstein-Hilbert action upon eliminating the spin connection via its
equation of motion. This amounts to expressing ${\cal R}^{ab}_{\mu \nu}$ in
terms of the Riemann tensor by inverting eq.\eqn{riecur} and then plugging
the result in eq.\eqn{qpala}. We have:
\begin{eqnarray}
&&q^{\Delta(a)} e^{\mu}_a e^{\nu}_b {\cal R}^{ab}_{\mu \nu}=-q^{\Delta(a)}
{\tt R}_{\mu \nu \rho}^{~~~~\tau} e^{\mu}_a e^{\nu}_b e^a_{\tau}
e^{b\rho}=\nonumber\\ &&=-q^{\Delta(b)} {\tt R}_{\mu \nu
\rho}^{~~~~\mu}e^{\nu}_b e^{b\rho}= {\tt R}_{\nu \mu \rho}^{~~~~\mu} {\tt
 g}^{\nu
\rho}={\tt R}~,\label{einhil1}
\end{eqnarray}
where we have made use of \eqn{prode}. Moreover we have:
\begin{eqnarray}
{\tt g} &\equiv& \det \parallel {\tt g}_{\mu \nu} \parallel =
\frac{1}{4!}\epsilon^{\mu_1 \mu_2 \mu_3
\mu_4} \epsilon^{\nu_1 \nu_2 \nu_3 \nu_4} {\tt g}_{\mu_1 \nu_1}
 {\tt g}_{\mu_2 \nu_2} {\tt g}_{\mu_3 \nu_3} {\tt g}_{\mu_4 \nu_4}=\nonumber\\
 &=&
 \frac{1}{4!} q^{\Delta(a_1)+\Delta(a_2)+\Delta(a_3)+\Delta(a_4)}
\epsilon^{\mu_1 \mu_2 \mu_3
\mu_4} \epsilon^{\nu_1 \nu_2 \nu_3 \nu_4}
 \eta_{a_1 b_1}\eta_{a_2 b_2}\eta_{a_3 b_3}\eta_{a_4 b_4}
 e^{a_1}_{\mu_1} e^{b_1}_{\nu_1}\cdots e^{a_4}_{\mu_4} e^{b_4}_{\nu_4}
 =\nonumber\\
 &=&
 \frac{1}{4!} q^{[\Delta(a_2)+\Delta(b_2)]+
 2[\Delta(a_3)+\Delta(b_3)]+
 3[\Delta(a_4)+\Delta(b_4)]}
\epsilon^{\mu_1 \mu_2 \mu_3
\mu_4} \epsilon^{\nu_1 \nu_2 \nu_3 \nu_4} \times \nonumber\\
&&\times \eta_{a_1 b_1}\eta_{a_2 b_2}\eta_{a_3 b_3}\eta_{a_4 b_4}
(e^{a_1}_{\mu_1}\cdots e^{a_4}_{\mu_4})
 (e^{b_1}_{\nu_1}\cdots e^{b_4}_{\nu_4})=\nonumber\\ &=&\frac{1}{4!}
\epsilon^{\mu_1 \mu_2 \mu_3
\mu_4} \epsilon^{\nu_1 \nu_2 \nu_3 \nu_4}
 \eta_{a_1 b_1}\eta_{a_2 b_2}\eta_{a_3 b_3}\eta_{a_4 b_4}
(e^{a_1}_{\mu_1}\cdots e^{a_4}_{\mu_4})
 (e^{b_1}_{\nu_1}\cdots e^{b_4}_{\nu_4})
 =\nonumber\\
 &=&\frac{1}{4!} q^{-12} \hat{\epsilon}_{a_1 a_2 a_3 a_4}
 \hat{\epsilon}^{a_1 a_2 a_3 a_4}{\tt e}^2=- q^{-6} \; {\tt
 e}^2~,\label{einhil2}
 \end{eqnarray}
where we made use of \eqn{eteqet}, \eqn{epsid} and \eqn{epep}. Putting
together
\eqn{einhil1} and \eqn{einhil2} we see that the q-Palatini action \eqn{qpala}
becomes equal to:
\be
{\cal S}=\frac{1}{2}\int_M d^4 x \, \sqrt{-{\tt g}}\; {\tt R}~,
\ee
which is the {\it undeformed} Einstein-Hilbert action. Since the components
of ${\tt g}_{\mu \nu}$ and its inverse all commute among-themselves, it is
clear that the equations of motion of the metric theory will be equal
to those of the undeformed Einstein's theory in vacuum. One can obtain the
same result starting directly from eq.\eqn{ffeq1} and using \eqn{riecur}.

Summarizing, the results of this Section show that if we just consider the
 theory constructed in terms of the space-time metric ${\tt g}_{\mu
\nu}$, our theory is completely equivalent to Einstein's theory.
 No trace of the q-structure existing in the tetrad formulation of the
theory can be found at the metric level. In the complete theory, which
includes the vierbeins and the spin connection, the space-time metric ${\tt
g}_{\mu\nu}$ can no longer be considered a c-number  (at each point in
space-time). Nevertheless, quantities like the Christoffel symbols, the
Riemann, Ricci and Einstein tensors, being in the center of the algebra
of functions on the space--time manifold,
can still be regarded as numbers. If one thus follows the view that in
classical general relativity all observables can be constructed out of the
latter quantities, we conclude that even the complete theory is {\it
physically} equivalent, at the classical level, to the ordinary Einstein's
theory.

\section{Inclusion of sources}
\setcounter{equation}{0}
Next let us introduce sources. We do this  by replacing the free field
equations (\ref{ffeq1}) and (\ref{ffeq2}) by
\begin{eqnarray}
\kappa \; \epsilon_{abcd} {\cal R}^{ab} \wedge e^{c}& =& \theta_d \;,
\cr \kappa \; q^{-\Delta(b)} \epsilon_{abcd} {\cal T}^{a}\wedge e^{b} &
=& \frac12 \Sigma_{cd} \;,  \label{feqs}
\end{eqnarray}
where $\theta_d$ and $\Sigma_{cd}$ are three-forms which are the analogues
of the stress-energy and spin densities. $\kappa$ is the gravitational
coupling constant. The second equation differs from the undeformed case by
factors of $q $. By substituting (\ref{feqs}) into  the Bianchi identities (
\ref{4dBi}) we get a simple set of consistency conditions for the sources
\begin{eqnarray}
d\theta_d&=&\omega_{dc}\wedge \theta^c +\kappa\epsilon_{abcd} {\cal R}^{ab}
\wedge {\cal T}^c \;,\cr d\Sigma_{ab}&=&q^{\Delta(a)} e_b\wedge \theta_a + {
\omega^c}_b \wedge \Sigma_{ac} -(a\rightleftharpoons b) \;.
\end{eqnarray}
These expressions also differ from the undeformed case by factors of $q$
. Moreover, we note that the three-forms $\theta_d$ and $\Sigma_{cd}$ are
not c-numbers. From the commutation relations (\ref{4dRTRT}), (\ref{4dRToe})
and (\ref{4dcrstc}), the source terms  should satisfy
\begin{eqnarray}
\theta_a \wedge\theta_b &=&-q^{\Delta(a)-\Delta(b)}\; \theta_b \wedge
\theta_{a} \;,\cr \theta_a \wedge\Sigma_{bc}
&=&-q^{2\Delta(a)-\Delta(b)-\Delta(c)}\; \Sigma_{bc}\wedge \theta_{a} \;,\cr
\Sigma_{ab} \wedge\Sigma_{cd}
&=&-q^{2\Delta(a)+2\Delta(b)-2\Delta(c)-2\Delta(d)} \; \Sigma_{cd}
\wedge\Sigma_{ab} \;,  \label{cpos}
\end{eqnarray}
where we used the identity (\ref{epsid}). Of course, the above exterior
products of three forms  vanish on a four dimensional manifold. We shall
interpret these equations as conditions on products of the components $
\theta_d^{\mu\nu\rho}$ and $\Sigma_{cd}^{\mu\nu\rho}$  of the three forms $
\theta_d$ and $\Sigma_{cd}$. That is the components satisfy
\begin{eqnarray}
\theta_a^{\mu\nu\rho} \;\theta_b^{\sigma\lambda\eta}
&=&q^{\Delta(a)-\Delta(b)}\; \theta_b ^{\sigma\lambda\eta} \; \theta_{a}
^{\mu\nu\rho} \;,\cr \theta_{a} ^{\mu\nu\rho}\;
\Sigma_{bc}^{\sigma\lambda\eta}
&=&q^{2\Delta(a)-\Delta(b)-\Delta(c)}\;\Sigma_{bc}^{\sigma\lambda\eta} \;
\theta_{a} ^{\mu\nu\rho} \;,\cr \Sigma_{ab} ^{\mu\nu\rho}\; \Sigma_{cd}
^{\sigma\lambda\eta} &=&q^{2\Delta(a)+2\Delta(b)-2\Delta(c)-2\Delta(d)} \;
\Sigma_{cd} ^{\sigma\lambda\eta}\; \Sigma_{ab}^{\mu\nu\rho} \;.
\end{eqnarray}
We can then replace $\theta_d$ and $\Sigma_{cd}$ in (\ref{cpos}) by their
corresponding dual one forms. The commutation relations involving the
source terms and the vierbeins or the spin-connections may be
derived from eqs.\eqn{feqs}, by consistency.

We next show how to recover the Einstein equations in the presence of
matter.  For this purpose we set the spin densities
 $\Sigma_{cd}$ equal to zero.  Then, assuming
inverse vierbeins to exist, we get that  the torsion vanishes. Next we rewrite
the first of equations (\ref{feqs}).   For this we start with
\be
q^{\Delta(b) -\Delta(c) -4\Delta(e)}
\hat{\epsilon}^{bcde} e^\lambda_b e^\rho_c e^\tau_e =-q^6
\epsilon^{\lambda\rho\sigma\tau} e^d_\sigma {\tt e}^{-1} \;,
\ee
which follows from  (\ref{id68}),  and multiply on the left with
$\epsilon_{fgdh}{\cal R}^{fg}_{\lambda\rho}$. Upon substituting in the
first of equations (\ref{feqs}), we then get
\be
q^{\Delta(b)-\Delta(c)-4\Delta(e)}
\hat{\epsilon}^{bcde} \epsilon_{fgdh}{\cal R}^{fg}_{\lambda\rho}
 e^\lambda_b e^\rho_c e^\tau_e =-\frac{q^6 }{3\kappa}\epsilon^{\lambda \rho
 \sigma \tau} \theta_{h
\lambda\rho\sigma}    {\tt e}^{-1} \;,
\ee
$ \theta_{h\lambda\rho\sigma}$ being the space time components of the three
form
$
\theta_{h}  $.    To evaluate the left hand side we can use (\ref{epsid}),
(\ref{ephaep})   and the identity
\be
-q^{-6}\hat{\epsilon}^{bcde} \epsilon_{fgdh}=
q^{-2\Delta(b)-\Delta(c)+\Delta(e)-3} (\delta^b_f
\delta^c_g\delta^e_h -  \delta^c_f
\delta^b_g\delta^e_h -
\delta^b_f\delta^e_g\delta^c_h
-\delta^b_h\delta^e_f\delta^c_g
+  \delta^b_h\delta^e_g\delta^c_f + \delta^b_g\delta^e_f\delta^c_h)
\;. \ee
We then multiply on the right  by $-1/4 \; q^{2\Delta(h)+ 2\Delta(a)+3}
e^a_\tau e^h_\mu$ to get the result
 \be
e^{\lambda}_c {\cal R}^{ac}_{\mu \lambda} - \frac{1}{2}e^{\lambda}_d
e^a_{\mu} e^{\rho}_e {\cal R}^{de}_{\lambda \rho}=\frac{1}{12 \kappa}
\Theta_\mu^a   \;,
\label{eqm2}
\ee
where
 \be
\Theta_\mu^a   =- q^{2\Delta(h)+2\Delta(a)+3}
\epsilon^{\lambda\rho\sigma\tau}   \theta_{h
\lambda\rho\sigma}    {\tt e}^{-1}e^a_\tau e^h_\mu\;.\label{eq81}
 \ee
Upon multiplying (\ref{eqm2}) and (\ref{eq81}) by $\eta_{af} e^f_{\nu}$ on
the right and, using eq.\eqn {riecur}, we can see that the left hand side
becomes equal to the Einstein tensor:
\begin{eqnarray}
 & &(e^{\lambda}_c {\cal R}^{ac}_{\mu\lambda} - \frac{1}{2}e^{\lambda}_d
 e^a_{\mu}
e^{\rho}_e {\cal R}^{de}_{\lambda \rho})\;\eta_{af} e^f_{\nu}= \nonumber \\
&
= & (-q^{\Delta(c)-\Delta(a)}{\tt R}_{\mu\lambda \rho}^{~~~~\sigma} e^a_{\sigma}
e^{\lambda}_c e^{\rho c}+ \frac{1}{2}q^{\Delta(e)-\Delta(a)} e^{\lambda}_d
e^d_{\sigma} e^a_{\mu} e^{\rho}_e e^{\tau e}{\tt R}_{\lambda \rho
\tau}^{~~~~\sigma})\; \eta_{af} e^f_{\nu}=\nonumber \\
&=&
-q^{-\Delta(a)}({\tt R}_{\mu\lambda \rho}^{~~~~\sigma}e^a_{\sigma} {\tt
 g}^{\lambda
\rho} -
\frac{1}{2}{\tt R}_{\lambda \rho \tau}^{~~~~\lambda}e^a_{\mu} {\tt g}^{\rho
 \tau} )\;
\eta_{af}
 e^f_{\nu}=
-{\tt R}_{\mu\lambda \rho}^{~~~~\sigma} {\tt g}^{\lambda \rho} {\tt g}_{\sigma
\nu} +
\frac{1}{2}{\tt R}_{\lambda \rho \tau}^{~~~~\lambda} {\tt g}^{\rho \tau} {\tt
 g}_{\mu
 \nu}=\nonumber\\
&=&{\tt R}_{\mu\lambda \nu \rho} {\tt g}^{\lambda \rho} -
\frac{1}{2}{\tt R}_{\rho
\lambda
\tau}^{~~~~\lambda} {\tt g}^{\rho \tau} {\tt g}_{\mu \nu}=
{\tt R}_{\mu \nu} - \frac{1}{2} {\tt g}_{\mu \nu} {\tt R}~,
\end{eqnarray}
where in the last line we made use of the second of eqs.\eqn{riem}. Thus
\eqn{eqm2} implies the usual Einstein's equation in the presence of matter:
\be
{\tt R}_{\mu \nu} - \frac{1}{2} {\tt g}_{\mu \nu} {\tt R}=
\frac{1}{12 \kappa} T_{\mu
\nu}~,
\ee
where $T_{\mu\nu}$ is the stress--energy tensor
\begin{eqnarray}
T_{\mu \nu}&=&\eta_{ab} \Theta^a_\mu e^b_\nu=-q^{2
\Delta(h)+2 \Delta(a)+3}\epsilon^{\lambda \rho \sigma
\tau}\eta_{ab} \theta_{h \lambda \rho \sigma} {\tt e}^{-1} e^a_{\tau} e^h_{\mu}
e^b_{\nu} =\nonumber\\ &=&-q^{
\Delta(h)+3}\epsilon^{\lambda \rho \sigma \tau}e^h_{\mu} \theta_{h \lambda \rho
\sigma} {\tt g}_{\tau \nu} {\tt e}^{-1}=
-q^{
\Delta(h)}\epsilon^{\lambda \rho \sigma \tau}e^h_{\mu} \theta_{h \lambda \rho
\sigma} {\tt g}_{\tau \nu} \frac{1}{\sqrt{-{\tt g}}}~.
\end{eqnarray}

We notice that the space-time components $T_{\mu \nu}$ of the matter
stress-energy tensor , having the same commutation properties as the
Einstein tensor, belong to the center of the algebra and thus can still be
interpreted as ordinary numbers. Thus, in our formalism, the properties of
matter, in the absence of sources for torsion, are left unaltered with
respect to the undeformed theory. Even in the presence of matter, the
metric version of our theory is then completely equivalent to Einstein's
theory, for all values of $q$.

\section{Hamiltonian formulation}
\setcounter{equation}{0}
We derive here the Hamiltonian formulation of Einstein's theory. We shall
do it by introducing a set of deformed Ashtekar's variables
\cite{ashtekar}. For simplicity we shall restrict to the source-free case.

The idea behind Ashtekar's variables is to use, instead of the
spin-connection components    $\omega^{ab}_{\mu}$,
 its {\rm self-dual} part $A^{ab}_{\mu}$
defined as:
\be
A^{ab}_{\mu}=\frac{1}{2} \omega^{ab}_{\mu} - \frac{i}{4} \epsilon_{cd}^{~~ab}
\omega^{cd}_{\mu}~\label{sda}
\ee
$A^{ab}_{\mu}$ satisfies the {\rm self-duality} property:
\be
i A^{cd}_{\mu}- \frac{1}{2} \epsilon_{ab}^{~~cd} A^{ab}_{\mu}=0~.\label{sdcond}
\ee
Its curvature $F^{ab}_{\mu \nu}$ is self-dual as well:
\be
i F^{cd}_{\mu \nu}- \frac{1}{2} \epsilon_{ab}^{~~cd} F^{ab}_{\mu
\nu}=0~,\label{sdcon2}
\ee
a property which plays a crucial role in rendering polynomial the
constraints of the theory. Now, our spin-connections are non-commuting
objects and thus eq.\eqn{sdcond} might be inconsistent. Fortunately this is
not the case: it can be checked that eq.\eqn{sda} only involves linear
combinations of the spin-connection with identical commutation properties
and thus $A^{ab}_{\mu}$ have the same commutation properties as those of
$\omega^{ab}_{\mu}$:
\begin{eqnarray}
A_\mu ^{ab} A_\nu ^{cd}=A_\nu ^{cd}A_\mu ^{ab}\;,\cr
e^\mu _a A_\nu ^{bc}=q^{-\Delta (b)-\Delta (c)}\;A_\nu ^{bc}e^\mu
_a\;,
\label{4dcra}
\end{eqnarray}
Similarly eq.\eqn{sdcond} involves components of $A^{ab}_{\mu}$ with
identical commutation properties and thus is consistent as well.

Following what is done in the undeformed case \cite{ashtekar}, we replace
the q-Palatini action \eqn{qpala} with:
\begin{equation}
{\cal S}=\int_M d^4 x \, q^{\Delta(a)} {\tt e} \, e^{\mu}_a e^{\nu}_b
F^{ab}_{\mu \nu}~.
\label{sdact}
\end{equation}
The above action is equivalent to \eqn{qpala}. To see this, we show that if
we solve first the equations of motion for $A^{ab}_{\mu}$ implied by
\eqn{sdact} (for a given arbitrary tetrad) and then plug the solution back
in the action, we get back \eqn{qpala}. Now, it is easy to verify that the
equations of motion for $A^{ab}_{\mu}$ imply, as in the undeformed case,
that $F^{ab}_{\mu \nu}$ be the self dual part of the curvature of the
spin-connection for the torsion-free connection determined by the tetrad.
Thus:
\be
F^{ab}_{\mu \nu}=\frac{1}{2} {\cal R}^{ab}_{\mu \nu} - \frac{i}{4}
\epsilon_{cd}^{~~ab}
{\cal R}^{cd}_{\mu \nu}~\label{sdcurv}
\ee
As in the undeformed case,
when we substitute  the right  hand side  of the above equation
 in \eqn{sdact} we see that it reduces to \eqn{qpala} (up to the irrelevant
constant factor $q^{-3}$)
 since:
\begin{eqnarray}
-q^{\Delta(a)} \epsilon_{cd}^{~~~ab} e^{\mu}_a e^{\nu}_b {\cal R}^{cd}_{\mu
\nu} &=& q^{\Delta(a)}\epsilon_{cd}^{~~~~ab} e^{\mu}_a e^{\nu}_b
{\tt R}_{\mu \nu \rho}^{~~~~\sigma} e^c_{\sigma} e^{\rho}_e \eta^{de}= q^{2
\Delta (a)+ \Delta(b)} \epsilon^{c a b d} e^{\mu}_a e^{\nu}_b e^{\rho}_d
{\tt R}_{\mu \nu \rho}^{~~~~\sigma} e_{\sigma c} =\nonumber\\ &=&q^{2
\Delta (a)+ \Delta(b)} \epsilon^{c a b d} e^{\mu}_a e^{\nu}_b e^{\rho}_d
{\tt R}_{[\mu
\nu \rho]}^{~~~~~\sigma} e_{\sigma c}
=0~,
\end{eqnarray}
where we used the Bianchi identity.
Thus we have the result of the equivalence of the self-dual action with
 Palatini's action.

We now sketch the canonical formalism for our q-deformed self-dual Palatini
action. In the canonical formalism one needs to split first the manifold
$M$ into a
 3-manifold $\Sigma$, playing the r\^ole of ``space", times a real
line associated with ``time", and then view the field equations as
giving the
``time" evolution of fields living on $\Sigma$. We perform the split of $M$
in the usual manner by foliating it by a collection of space-like surfaces
coinciding with  $t=$constant level surfaces for some (real) function
$t$. The ``time" variable is then introduced by means of a time-like real
vector field $t^{\mu}$, whose integral curves intersect every leaf of the
foliation at a unique point. In this way we can identify all the leaves
with a standard space-like surface $\Sigma$, which is our ``space".
$t^{\mu}$ is normalized such that
\be
t^{\mu} \; \del_{\mu}t=1 \label{normt}
\ee
and the Lie derivative along $t^{\mu}$ will play the r\^ole of the
``time derivative".

Next let $n_{\mu}$ be the unit covector normal to the leaves of the
foliations. It must be proportional to the gradient of $t$:
\be
n_{\mu} = -{\tt N}\; \del_{\mu} t~,\label{eqn59}
\ee
where $\tt N$ is the analogue of the lapse function. In our q-calculus $\tt
N$ (and thus $n_{\mu}$) is not an ordinary number. Rather it fulfills the
commutation relations:
$$
{\tt N}\; e^{\mu}_a = q^{-\Delta(a)} e^{\mu}_a \;{\tt N}~,~~~{\tt N}\; {\tt
e}
= {\tt e}\;{\tt N}~,
$$
\begin{equation}
{\tt N} \;\omega_{\mu}^{ab} = q^{\Delta(a)+ \Delta(b)} \omega_{\mu}^{ab}\;
{\tt N}~.
\end{equation}
The commutation properties of ${\tt N}^{-1}$ easily follow from the above
formulae. ${\tt N}$ is chosen such that $n_{\mu}$ is a unit vector:
\begin{equation}
n^{\mu} n_{\mu} \equiv {\tt g}^{\mu \nu} n_{\mu} n_{\nu}= {\tt g}^{\mu \nu}
\;{\tt N}^2 \; \del_{\mu} t \del_{\nu}t=-1~.
\end{equation}
The commutation properties of $\tt N$ render the above equation consistent,
as its left hand side commutes with everything.
(Had $\tt N$ been an ordinary c-number
this would have not been true.)
 We now decompose $t^{\mu}$ into normal
and tangential components. It follows from \eqn{normt} that:
\be
t^{\mu}= n^{\mu}\; {\tt N} + N^{\mu}~,\label{decom}
\ee
where $N^{\mu}$ (the shift)
is an ordinary vector such that $N^{\mu} n_{\mu}=0$.  [This
condition is consistent with (\ref{normt}) and (\ref{eqn59})].  Notice
that even though niether
 $\tt N$ nor $n^{\mu}$ are commuting objects their
product, being in the center of the algebra, can be considered to be so,
which is the only thing we need in order to write \eqn{decom}.

In order to perform an analogous 3+1 split of the fields, we now
introduce the projection operator $q^{\mu}_{\nu}$:
\be
q^{\mu}_{\nu} \equiv \delta^{\mu}_{\nu} + n_{\nu} n^{\mu}~.
\ee
(Notice that the components of $q^{\mu}_{\nu}$ commute with everything.)
With its help we decompose the tetrads $e_a^{\mu}$ into components
normal and tangential to $\Sigma$:
\be
e^{\mu}_a=E^{\mu}_a - n_a n^{\mu}~,\label{split}
\ee
where
\be
E^{\mu}_a= q^{\mu}_{\nu} \; e^{\nu}_a~,
\ee
\be
n_a \equiv e^{\mu}_a \; n_{\mu}~.
\ee
Even if it is written as a four-dimensional field, $E^{\mu}_a$ has to be
thought as a field living on $\Sigma$, because contracting it with any
vector normal to $\Sigma$ gives zero. In the following
  we shall stress the difference
among three dimensional fields like $E^{\mu}_a$ and four dimensional ones
by writing an arrow over the former.

 Upon substituting eq.\eqn{split} in   \eqn{sdact} we get:
\begin{eqnarray}
{\cal S} & = & \int_M d^4 x \, q^{\Delta(a)} {\tt e} \, e^{\mu}_a e^{\nu}_b
F^{ab}_{\mu \nu}=
\int_M d^4 x \,
(q^{\Delta(a)} {\tt e} \, \vec{E}^{\mu}_a \vec{E}^{\nu}_b F^{ab}_{\mu
\nu}~- 2 q^{\Delta(a)} {\tt e} \, \vec{E}^{\mu}_a n_b n^{\nu} F^{ab}_{\mu
\nu}~)=\nonumber \\ & = &
\int_M d^4 x \, (q^{\Delta(a)} {\tt e} \, \vec{E}^{\mu}_a \vec{E}^{\nu}_b
 F^{ab}_{\mu
\nu}~ +i q^{\Delta(a)} {\tt e} \, \vec{E}^{\mu}_a n_b n^{\nu} \epsilon
^{ab}_{~~cd} F^{cd}_{\mu \nu}~)~,
\end{eqnarray}
where we have made use of eq.\eqn{sdcon2} in the last passage. Upon
introducing the q-triad field $\tilde{\vec{E}}^{\mu}_a = {\tt N}^{-1}{\tt
e}
\vec{E}^{\mu}_a$ and the quantity $\tilde{\tt N}= {\tt N}^2 {\tt e}^{-1}$
(we shall write a tilde on field densities), we rewrite the last line of
the above expression as
\begin{eqnarray}
 & &
\int_M d^4 x \,
[q^{-2\Delta(a)} \tilde{{\tt N}} \, \tilde{\vec{E}}^{\mu}_a
\tilde{\vec{E}}^{\nu}_b F^{ab}_{\mu \nu}~ +i q^{-\Delta(b)} \,
 \tilde{\vec{E}}^{\mu}_a
n_b (t^{\nu}-N^{\nu}) \epsilon ^{ab}_ {~~cd} F^{cd}_{\mu \nu}]~=\nonumber\\
&=&
\int_M d^4 x \, [q^{-2\Delta(a)} \tilde{{\tt N}} \, \tilde{\vec{E}}^{\mu}_a
\tilde{\vec{E}}^{\nu}_b F^{ab}_{\mu \nu}~ -i q^{-\Delta(b)} \,
 \tilde{\vec{E}}^{\mu}_a
n_b  \epsilon ^{ab}_{~~cd} ({\cal L}_t A^{cd}_{\mu} - D_{\mu}(A^{cd}_{\nu}
t^{\nu}) - N^{\nu} F^{cd}_{\nu
\mu})]~=\nonumber\\
&=&\int_M d^4 x
\, [-i q^{-\Delta(b)} \, \tilde{\vec{E}}^{\mu}_a n_b  \epsilon ^{ab}_{~~cd}
 {\cal
L}_t A^{cd}_{\mu}+ i q^{-\Delta(b)}
 \, \tilde{\vec{E}}^{\mu}_a n_b  \epsilon ^{ab}_{~~cd}
N^{\nu}F^{cd}_{\nu \mu}~+ \nonumber\\ &-& i
q^{-\Delta(b)}D_{\mu}(\tilde{\vec{E}}^{\mu}_a n_b
\epsilon ^{ab}_{~~cd})
(A^{cd}_{\nu} t^{\nu})+ q^{-2\Delta(a)} \tilde{{\tt N}} \,
\tilde{\vec{E}}^{\mu}_a \tilde{\vec{E}}^{\nu}_b F^{ab}_{\mu \nu}]~.\label{leg}
\end{eqnarray}
In deriving the above equation we have used the identity  $t^{\nu}(F_{\mu
\nu}^{ab})=-{\cal L}_t(A_{\mu}^{ab})+ D_{\mu}(A^{cd}_{\nu}
t^{\nu})$  in the second line,
where ${\cal L}_t$ is the Lie derivative along the vector field
$t^{\mu}$ and $ D_{\mu}$ is the covariant derivative relative to
$A_{\mu}^{ab}$. Notice also that in
$D^{}_{\mu}(\tilde{\vec{E}}^{\mu}_a n_b \epsilon ^{ab}_{~~cd})$,
 the field $A^{ab}_{\mu}$ has to be written on the
 $right$, while in  $D_{\mu}(A^{cd}_{\nu}  t^{\nu})$,
 $A^{ab}_{\mu}$ has to be written on the $left$.

Now we observe that all fields in eq.\eqn{leg} are written in a 3+1 form.
First consider the terms containing $F^{ab}_{\mu
\nu}$.  Since $F^{ab}_{\mu \nu}$ always appears contracted with vectors
lying in $\Sigma$, we can replace it with the curvature of the pull-back
$\vec{A}^{ab}_{\mu}$ of $A^{ab}_{\mu}$ to $\Sigma$, which we denote by
$\vec{F}^{ab}_{\mu \nu}$.
($\vec{F}^{ab}_{\mu \nu}$ is the analogue of the  magnetic field.)
 Also, since $q^{\mu}_{\nu} {\cal L}_t q^{\rho}_{\mu}=0$,
we can replace ${\cal L}_t A^{cd}_{\mu}$ with ${\cal L}_t
\vec{A}^{cd}_{\mu}$ in the first term of the result of
\eqn{leg}. Finally, in the term containing $D_{\mu}(\tilde{\vec{E}}^{\mu}_a
n_b
\epsilon ^{ab}_{~~cd})$ we can replace $D_{\mu}$ by its projection
$\vec{D}_{\mu}$ onto $\Sigma$, obtained by replacing in it $A^{ab}_{\mu}$
with $\vec{A}^{ab}_{\mu}$. As for $ A^{cd}_{\nu} t^{\nu}$, it plays the
r\^ole of the ``time "component of the connection $A^{cd}_{\mu}$. Thus we
can rewrite eq.\eqn{leg} in the 3+1 form:
\begin{eqnarray}
{\cal S}&=&\int_M d^4 x
\, [-i q^{-\Delta(b)} \, \tilde{\vec{E}}^{\mu}_a n_b  \epsilon ^{ab}_{~~cd}
\dot{\vec{A}}^{cd}_{\mu}+ i q^{-\Delta(b)}
 \, \tilde{\vec{E}}^{\mu}_a n_b  \epsilon ^{ab}_{~~cd}
N^{\nu}\vec{F}^{cd}_{\nu \mu}~+ \nonumber\\ &-& i
q^{-\Delta(b)}\vec{D}_{\mu}(\tilde{\vec{E}}^{\mu}_a n_b
\epsilon ^{ab}_{~~cd})
(A^{cd}_{\nu} t^{\nu})+ q^{-2\Delta(a)} \tilde{{\tt N}} \,
\tilde{\vec{E}}^{\mu}_a \tilde{\vec{E}}^{\nu}_b \vec{F}^{ab}_{\mu
 \nu}]~,\label{pqdot}
\end{eqnarray}
where we have adopted the notation $\dot{\vec{A}}^{ab}_{\mu}$ for ${\cal
L}_t \vec{A}^{ab}_{\mu}$.

In eq.\eqn{pqdot} the action is  written in
the form $\int dt(P \dot{Q} - H)$ and we thus can read off the canonical
coordinates. The r\^ole of $Q$ is played by the pull-back to $\Sigma$,
$\vec{A}_{\mu}^{ab}$, of the self-dual connection $A_{\mu}^{ab}$, while its
conjugate momentum $\tilde{\vec{\Pi}}^{\mu}_{ab}$ is:
\be
\tilde{\vec{\Pi}}^{\mu}_{cd}=  q^{-\Delta(d)}
 \, \tilde{\vec{E}}^{\mu}_c n_d- q^{-\Delta(c)}
 \, \tilde{\vec{E}}^{\mu}_d n_c -\frac{i}{2} q^{-\Delta(b)}
 \, \tilde{\vec{E}}^{\mu}_a n_b  \epsilon ^{ab}_{~~cd}~.
\ee
Thus the non-vanishing Poisson brackets are:
\begin{eqnarray}
\{ \vec{A}_{\mu}^{ab}(x) ,\tilde{\vec{\Pi}}^{\nu}_{cd}(y)\}&=&-q^{-2(\Delta(a) +
 2
\Delta(b))}
\{ \tilde{\vec{\Pi}}^{\nu}_{cd}(y),\vec{A}_{\mu}^{ab}(x) \}=\nonumber\\
&=&\frac{1}{2}
\delta^{\nu}_{\mu}\;
 [\delta^a_{ [\; c} \delta^b_{d]}-\frac{i}{2}\epsilon ^{ab}_{~~cd}\;]
 \; \delta^3(x,y) ~.
\end{eqnarray}
Notice that our Poisson brackets are not skewsymmetric, as a consequence of
the following non-trivial commutation properties of $\vec{A}_{\mu}^{ab}$
and $\tilde{\vec{\Pi}}^{\mu}_{ab}$ :
\begin{eqnarray}
\tilde{\vec{\Pi}}^{\mu}_{ab}\; \tilde{\vec{\Pi}}^{\nu}_{cd}&=&
 q^{2(\Delta(a)+\Delta(b)
-\Delta(c)-\Delta(d))} \;
\tilde{\vec{\Pi}}^{\nu}_{cd}\;\tilde{\vec{\Pi}}^{\mu}_{ab}  ~,\nonumber\\
 \tilde{\vec{\Pi}}^{\mu}_{ab}\; \vec{A}^{cd}_{\nu}&=&q^{2(\Delta(c)+\Delta(d))}
 \;
 \vec{A}^{cd}_{\nu}\; \tilde{\vec{\Pi}}^{\mu}_{ab}~.
 \end{eqnarray}
We can now rewrite the action in terms of the canonical variables as:
\be
{\cal S}=\int_M d^4 x\; \left\{ \tilde{\vec{\Pi}}^{\mu}_{ab}
\dot{\vec{A}}^{ab}_{\mu} + ( \vec{D}_{\mu}\tilde{\vec{\Pi}}^{\mu}_{~cd})
 (A^{cd}_{\nu}
t^{\nu})+
\tilde{{\tt N}} \; q^{-2\Delta(a)+2 \Delta(d)} \eta^{cd}
\tilde{\vec{\Pi}}^{\mu}_{ac}\tilde{\vec{\Pi}}^{\nu}_{db}  \vec{F}^{ab}_{\mu
 \nu}-
N^{\mu} \tilde{\vec{\Pi}}^{\nu}_{ab} \vec{F}^{ab}_{\mu
\nu}\right\}~.\label{actcan}
 \ee
Since eq.\eqn{actcan} contains no ``time derivatives" , namely
Lie derivatives with respect to $t^{\mu}$, of the fields $\tilde{{\tt N}}$,
$N^{\mu}$ and $ A^{cd}_{\nu} t^{\nu}$, we conclude that they are Lagrange
multipliers for the constraints:
\begin{eqnarray}
&&q^{-2\Delta(a)+2\Delta(d)}\eta^{cd}
\tilde{\vec{\Pi}}^{\mu}_{ac}\tilde{\vec{\Pi}}^{\nu}_{db}\vec{F}^{ab}_{\mu \nu}
 \approx 0~,\nonumber\\
&&
\vec{D}_{\mu}\tilde{\vec{\Pi}}^{\mu}_{~cd}\equiv
\partial_{\mu}\tilde{\vec{\Pi}}^{\mu}_{cd}
 +\tilde{\vec{\Pi}}^{\mu}_{ed}\vec{A}^{~~~e}_{\mu
c}\approx~0~,\nonumber\\ &&
\tilde{\vec{\Pi}}^{\nu}_{ab} \vec{F}^{ab}_{\nu \mu}\approx 0.\label{constr}
\end{eqnarray}
It is clear from the above formulae that our deformed canonical formalism
reduces to the undeformed one, as it is presented for example in the book
\cite{ashtekar}, for $q\rightarrow 1$.

\section{Concluding remarks}
\setcounter{equation}{0}
In this paper we have shown how to build a theoretical scheme in which
Einstein's classical theory of general relativity enters as the invariant
kernel common to a one-parameter family of theories. The main theoretical
tool has been the construction of gauge theories based on
a $q$-deformed rather than an ordinary Lie group \cite{cast1}. This
was  made possible by extending, by means of purely algebraic
techniques, the ordinary bicovariant (right/left covariant) calculus on
group manifolds to the deformed case \cite{cast96}. This procedure has been
successfully carried out for the Poincar\'e group \cite{castcom,cast2}.

We have seen in section 6 how one can extend Ashtekar's approach to build a
canonical formalism for any value of $q$. Apart from a minor change in one
of the constraints, the noncommutative nature of the conjugate variables
is reflected in the structure of the Poisson brackets which are no longer
skewsymmetric. We do not yet know  the consequences of
having an entire family of Hamiltonian formalisms for
general relativity  at our disposal.  We recall that there exists
 an analogous
 occurrence of different Hamiltonian structures for two dimensional
integrable models(although there the fields are c-numbers)\cite{rsv}.

Apart from its formal aspects, the physical content of our theoretical
scheme is bound to its quantization. One is then faced with the new
technical problem of quantizing a canonical theory of noncommuting
canonical variables. Clearly before attempting to quantize the system
presented here,
 initial efforts should be devoted to quantizing analogous
$q$-deformed systems in classical mechanics, probably in terms of a
path-integral formulation. New developments \cite{cast3} in integration
techniques on the quantum plane seem to open the way in this direction. The
first step toward quantum gravity could then be taken by tackling the
problem in 2+1 dimensions where a $q-$deformed Chern-Simons formulation of
gravity is available \cite{bmsv}. As the topological nature of the theory
is preserved by the deformation, one expects to be able to still
characterize expectation values of physical quantities, related to
noncontractible loops of the space-manifold, in terms of quantized knots
invariants.

It is  our hope that the existence of a one-parameter
family of $q$-deformed formulations of general relativity, each endowed
with an Hamiltonian structure, will shed some light on quantum gravity in
physical space-time dimensions. One could imagine a scenario where $q$
plays the r\^ole of a regularization parameter, with the advantage that the
correct classical limit is obtained by taking the  limit of vanishing
Planck's constant for any value of $q$.

\section{Acknowledgments}

We thank G. Marmo for the stimulating discussions while the manuscript was in
preparation. We also thank L. Castellani for discussing with us the results of
 \cite{cast3}. R. M. acknowledges E.E.C. contract FMRXCT96-0045 and a grant
of Italian Ministry of Education and Scientific Research (MURST 40\%).
A.S. was supported in part by the U.S. Department of Energy under
contract number DE-FG05-84ER40141.

\end{document}